\documentclass[superscriptaddress,twocolumn,prl,amsmath,amssymb,floatfix]{revtex4-2}
\usepackage{graphicx}
\usepackage{dcolumn}
\usepackage{bm}

\def\R{\mathcal{R}}
\def\Rhat{\hat{\mathcal{R}}}
\def\rhohat{\hat{\rho}}
\def\drho{\Delta \rho}
\def\sgn{\mathrm{sgn}}
\def\W{\bm{W}}
\def\Win{\bm{W}_\mathrm{in}}
\def\Wout{\bm{W}_\mathrm{out}}

\def\wout{\bm{w}_\mathrm{out}}

\def\Poincare{Poincar\'{e}~}

\renewcommand{\eqref}[1]{(Eq.~\ref{#1})}
\newcommand{\mean}[1]{\langle #1 \rangle}

\newcommand{\three}{Sec.~I\hspace{-1.2pt}I\hspace{-1.2pt}I}
\newcommand{\four}{Sec.~I\hspace{-1.2pt}V}
\newcommand{\five}{Sec.~V}
\newcommand{\six}{Sec.~V\hspace{-1.2pt}I}
\newcommand{\seven}{Sec.~V\hspace{-1.2pt}I\hspace{-1.2pt}I}
\newcommand{\eight}{Sec.~V\hspace{-1.2pt}I\hspace{-1.2pt}I\hspace{-1.2pt}I}

\newcommand{\FigureOne}{%
    \begin{figure}[t!]
        \begin{center}
         \includegraphics[scale=1.0]{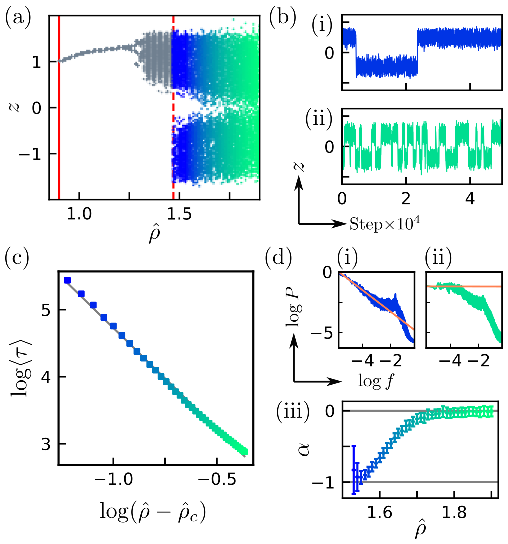}
         \caption{
            Bifurcation of a binary-feedback system induced by  $\rhohat$. 
            (a) Bifurcation diagram plotting the extrema of $z = \wout^{\top} \bm{x}$. This system is trained at $\rho=0.9$ (the solid line), and the critical point is $\rhohat_c \approx 1.47$ (the dashed line).
            (b) $z$ trajectories for (i) $\rhohat=1.56$ and (ii) $\rhohat=1.80$.
            (c, d) Statistical analysis of $z$ time series over $10^6$ steps with $50$ different initial states for each $\rhohat > \rhohat_c$.
            (c) Average interval $\mean{\tau}$ of sign changes.
            (d) (i,~ii) PSD at $\rhohat=1.56$ and $\rhohat=1.80$, with fitted lines for $f < 10^{-4}$, showing slopes $\alpha$ of $-0.88$ and $0.00$, respectively. (iii) Slope $\alpha$ at each $\rhohat$. 
        }
        \label{fig:fp09}
        \end{center}
    \end{figure}
}

\newcommand{\FigureTwo}{%
    \begin{figure}[t!]
        \begin{center}
         \includegraphics[scale=1.0]{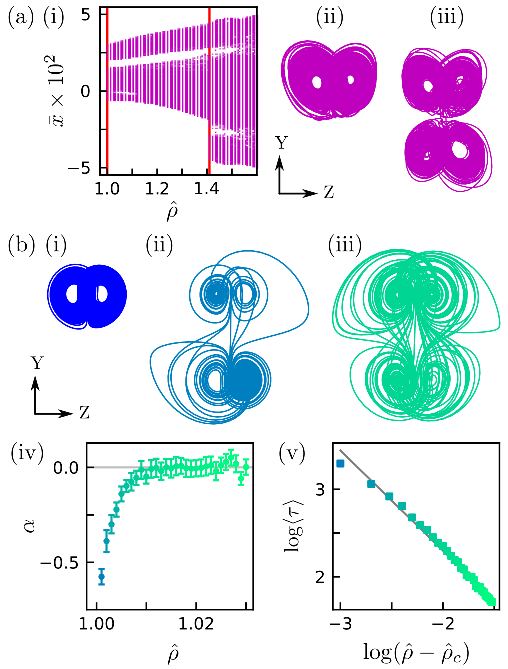}
         \caption{
            (a) Bifurcation of the switching Lorenz-driven system trained at $\rho = 1.0$. 
            The critical point is $\rhohat_c \approx 1.41$.
            (i) Bifurcation diagram plotting the extrema of the mean internal activity $\bar{x}$. 
            (ii,~iii) Output at $\rhohat=1.4, 1.5$.
            (b) Bifurcation of the Lorenz-embedded system. 
            (i--iii) Output at $\rhohat = 1.0, 1.001,$ and $1.01$, respectively.
            The critical point is $\rhohat_c \approx \rho = 1.0$.
            (iv,~v) Statistical analysis of the $Z$ component time series over ${10}^5$ steps from $20$ different initial conditions;
            slope $\alpha$ of the PSD for $f < {10}^{-4}$ on a log--log scale and the avarage interval $\mean{\tau}$ of the sign changes at each $\rhohat$.
        }
        \label{fig:Lorenz}
        \end{center}
    \end{figure}
}

\newcommand{\FigureThree}{%
    \begin{figure}[t!]
        \begin{center}
            \includegraphics[scale=1.0]{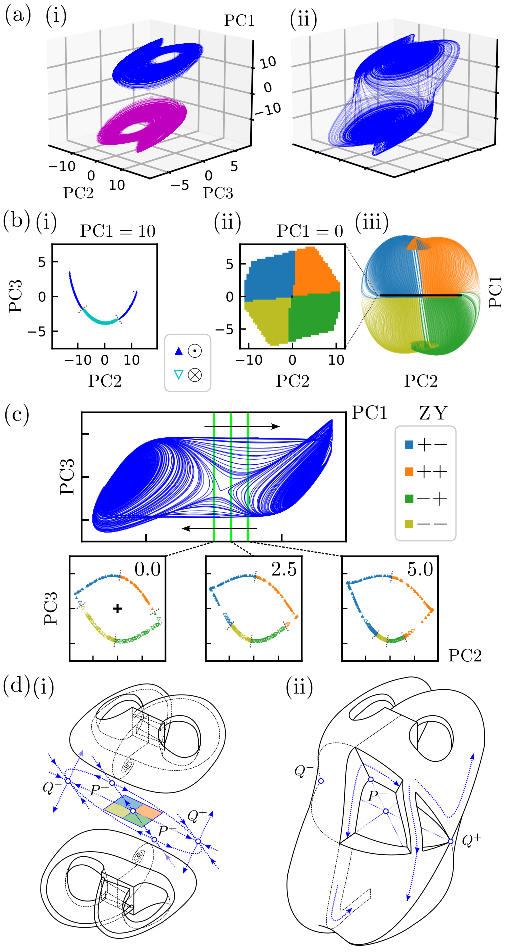}
            \caption{
            AMC in the Lorenz-embedded model.
            (a) Orbit in the PC space at (i) $\drho=-10^{-7}$ with bistability and (ii) at $\drho=10^{-3}$ with intermitency. 
            (b) For $\drho=-10^{-7}$: (i) \Poincare section at $\mathrm{PC}2>0$; (ii) color-coded initial points on $\mathrm{PC}1=0$ ($\mathrm{max}|x_i|<1$) based on the sign of $(Z, Y)$ when reaching $|Z|=27$; (iii) sample trajectories (200 steps).  
            (c) For $\drho=10^{-3}$: trajectories with $\mathrm{PC}2>0$ in the PC3--PC1 plane and \Poincare sections at $\mathrm{PC}1\in{0.0,~2.5,~5.0}$, with colored points as in B-ii.
            (d) Geometric flow model (i) before and (ii) after the crisis, based on~\cite{guckenheimer1979structural}. The colored plane in (i) corresponds to (b)-(ii).
        }
        \label{fig:Lorenz_crisis}
        \end{center}
    \end{figure}
}

\newcommand{\FigureFour}{%
    \begin{figure}[t!]
        \begin{center}
            \includegraphics[scale=1.0]{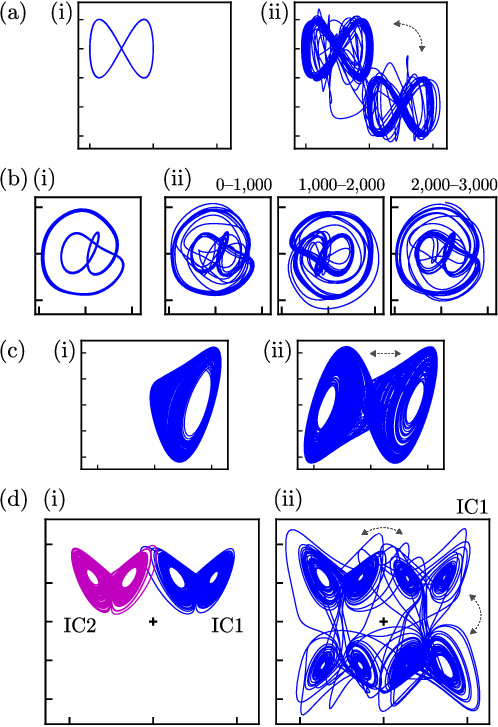}
            \caption{
            (a) AMC of Lissajous curve--shaped chaos.
            (i) Chaotic attractor designed at $\rho = 1.4440$.
            (ii) Behavior at $\rhohat = 1.4445$.
            (b) AMC of ``at"-shaped chaos.
            (i) Chaotic attractor designed at $\rho = 1.300$.
            (ii) Behavior at $\rhohat = 1.305$: Output in intervals $t \in [0, 1000], [1000, 2000],$ and $[2000, 3000]$.
            (c) AMC of the RC system trained on Chua's circuit spiral attractor.
            (i) Trajectory at $\rhohat = \rho = 1.5$.
            (ii) Double-scroll attractor at $\rhohat = 1.55$.
            (d) Crisis in the asymmetric case.
            (i) Two trajectories at $\rhohat = \rho = 1.3$ with different initial conditions (IC1 and IC2) of the MFRC system trained on a pair of Lorenz attractors.
            (ii) A trajectory at $\rho = 1.301$ from IC1.
        }
        \label{fig:example}
        \end{center}
    \end{figure}
}

\begin{document}
\title{
    Attractor-merging Crises and Intermittency in Reservoir Computing
}

\author{Tempei Kabayama}
\affiliation{
 Graduate School of Information Science and Technology, University of Tokyo, Bunkyo-ku, Tokyo 113-0033, Japan
}
\author{Motomasa Komuro}
\affiliation{
 International Research Center for Neurointelligence, University of Tokyo Institutes for Advanced Study, University of Tokyo, Bunkyo-ku, Tokyo 113-0033, Japan
}
\author{Yasuo Kuniyoshi}
\affiliation{
 Graduate School of Information Science and Technology, University of Tokyo, Bunkyo-ku, Tokyo 113-0033, Japan
}
\author{Kazuyuki Aihara}
\affiliation{
 International Research Center for Neurointelligence, University of Tokyo Institutes for Advanced Study, University of Tokyo, Bunkyo-ku, Tokyo 113-0033, Japan
}
\author{Kohei Nakajima}
\affiliation{
 Graduate School of Information Science and Technology, University of Tokyo, Bunkyo-ku, Tokyo 113-0033, Japan
}
\affiliation{
 International Research Center for Neurointelligence, University of Tokyo Institutes for Advanced Study, University of Tokyo, Bunkyo-ku, Tokyo 113-0033, Japan
}

\date{\today}

\begin{abstract}
    Reservoir computing can embed attractors into random neural networks (RNNs), generating a ``mirror'' of a target attractor because of its inherent symmetrical constraints. In these RNNs, we report that an attractor-merging crisis accompanied by intermittency emerges simply by adjusting the global parameter. We further reveal its underlying mechanism through a detailed analysis of the phase-space structure and demonstrate that this bifurcation scenario is intrinsic to a general class of RNNs, independent of training data.
\end{abstract}

\maketitle

Neural networks (NNs) are crucial for machine learning and modeling. However, employing them requires careful recognition of their intrinsic properties that should be appropriately harnessed.
Transient and intermittent dynamics arise not only in biological neural activity~\cite{Aihara1987, freeman1987simulation, yonekura2025reservoir} but also in artificial intelligence~\cite{inoue2022transient, storm2024finite}.
Moreover, such dynamics have been applied in engineering fields, including information processing and optimization~\cite{tokuda1997global, liu2024exploiting}.
This study mainly focuses on intermittency in NNs.

Crises serve as a mechanism for the emergence of intermittency~\cite{grebogi1983crises, grebogi1987critical}.
This is a typical bifurcation characterized by a discontinuous change of a chaotic attractor due to contact with an unstable periodic orbit.
In symmetric systems, multiple attractors can simultaneously touch an unstable orbit on their basin boundary, causing an attractor-merging crisis (AMC).
The system then exhibits crisis-induced intermittency, alternating between staying within and escaping the attractor's remnants, which is called ``attractor ruin''~\cite{namikawa2005chaotic}.

In this paper, we discuss crises and intermittency observed in random NNs (RNNs) within the machine learning framework called reservoir computing (RC)~\cite{jaeger2001echo, maass2002real, nakajima2021reservoir}.
In RC, an RNN  called ``reservoir" is employed.
This has a $D$-dimensional input $\bm{u}_k$ and an $N$-dimensional state $\bm{x}_k$, evolving as
\begin{equation}
    \bm{x}_{k+1} = \R(\bm{x}_k, \bm{u}_k).
    \label{eq:reservoir}
\end{equation}
By mapping $\bm{x}_k$ to a target output $\bm{y}_k$, RC conducts various signal processing of $\bm{u}_k$. 
The output $\bm{z}_k$ is given by
\begin{equation}
    \bm{z}_{k}=\phi(\Wout^\top\psi(\bm{x}_k)),
    \label{eq:output}
\end{equation}
where the functions $\psi$ and $\phi$ are nonlinear functions, including the identity map.

When $\bm{y}_k = \bm{u}_k$, RC performs a one-step-ahead prediction of the input, and feeding back the predicted value $\bm{z}_k$ as an alternative input constructs an autonomous system:
\begin{equation}
    \bm{x}_{k+1}=\R(\bm{x}_{k}, \bm{z}_{k})
                =\Rhat(\bm{x}_{k}).
    \label{eq:closedloop}
\end{equation}
This closed-loop configuration enables the attractor reconstruction from a finite-length time series~\cite{lu2018attractor, hart2020embedding}.

Several previous studies have approached transient dynamics in RC. \citet{kong2021emergence, kong2021machine} utilized parameter-aware RC~\cite{kim2021teaching}, which learns multiple samples of a parametric family to infer bifurcation structure. 
They estimated crises from pre-crisis time series and reproduced plausible transients. 
\citet{flynn2023seeing} reported that in multifunctional RC (MFRC), which embeds multiple training data to achieve multistability, a certain parameter region leads to the collapse of multistability and the emergence of intermittency.
\citet{flynn2023theory} further showed that even well-trained MFRC models can exhibit similar behavior when parameters are adjusted post hoc, and \citet{flynn2024exploring} analyzed its underlying mechanisms in specific settings.

In contrast, we focus on cases where both multistability and transient responses emerge intrinsically---independent of learning---and approach the general properties of RNNs.
In reservoir \eqref{eq:reservoir} and readout \eqref{eq:output}, if the symmetries 
$\R(-\bm{x}, -\bm{u})=-\R(\bm{x}, \bm{u})$ and $\phi(-\bm{y})=-\phi(\bm{y})$
hold, the closed-loop system \eqref{eq:closedloop} exhibits the symmetry $\Rhat(-\bm{x})=-\Rhat(\bm{x})$.
Therefore, when an attractor $\Lambda$ is embedded, it will also have the ``mirror'' attractor $-\Lambda$, and the system is multistable if $\Lambda\neq-\Lambda$.

As for the reservoir $\R$, we consider the following echo state network (ESN)~\cite{jaeger2007optimization}:
\begin{equation}
    \R(\bm{x}, \bm{u}) = (1-a)\bm{x} + a \tanh(\rho \W \bm{x} + \sigma \Win \bm{u}), \label{eq:lesn}
\end{equation}
Where $\W \in \mathbb{R}^{N \times N}$ is a random matrix where the proportion of nonzero elements is $p$, and the values are drawn from a normal distribution and then scaled to have a unity spectral radius so that $\rho$ is the spectral radius of the internal matrix $\rho \bm{W}$.  
Each element of the random matrix $\Win \in \mathbb{R}^{N \times D}$ is drawn from a uniform distribution on $[-1, 1]$, and $\sigma$ represents the input intensity. 
The parameter $a$ is the leaky rate.  
Unless otherwise specified, we fix the parameters as $(N, p, a) = (1000, 0.01, 0.5)$ in this paper.

The parameter $\rho$ is crucial for the reservoir's convergence properties and memory capacity~\cite{sompolinsky1988chaos, cessac2007neuron, jaeger2007optimization, haruna2019optimal}.  
When $\sigma$ is fixed, the conditional Lyapunov exponent of the input-driven reservoir increases with $\rho$~\cite{verstraeten2007experimental, lu2018attractor}.  
In particular, when the input is constant, the attractor of $\bm{x}_{k+1}=\R(\bm{x}_{k}, c)$ transitions from a fixed point to chaos, given a sufficiently large $N$ (see Supplementary Material (SM) \four).

Owing to the symmetry of ESN, the attractors of $\R(\bm{x}, +c)$ and $\R(\bm{x}, -c)$ are symmetric about the origin.
This property enables the construction of the following binary-feedback system:
\begin{equation}
    \bm{x}_{k+1} = \R(\bm{x}_{k}, \sgn(\wout^\top \bm{x}_{k})).
    \label{eq:binaryfeedback}
\end{equation}
The vector $\wout\in\mathbb{R}^{N}$ serves as a linear classifier that distinguishes between the attractors $\Lambda^{\pm}$ of $\R(\bm{x}, \pm1)$ for a given $\rho$, and it is obtained using a support vector machine.  
If $\Lambda^{\pm}$ are linearly separable, the system \eqref{eq:binaryfeedback} becomes bistable. 
In particular, if the attractors are fixed points $\pm\bm{p}$, they are clearly linearly separable, and $\wout=\bm{p}/\|\bm{p}\|^2$.

When $(\rho, \sigma)=(0.9, 0.2)$, the obtained system has two fixed points at $z=\bm{w}_{\mathrm{out}}^{\top} \bm{x}=\pm1$. The system can be destabilized by fixing $\wout$ and increasing $\rho$. To distinguish this bifurcation parameter from the hyperparameter in training, we denote it as $\rhohat$.
Figure~\ref{fig:fp09}(a) shows that as $\rhohat$ increases, fixed points transition to periodic orbits and then to chaos around $\rhohat = 1.3$, resulting in two coexisting chaotic attractors.
A crisis occurs at $\rhohat=\rhohat_c\simeq1.47$, where the attractors merge.
This is likely triggered by the expansion of both the attractor and the preimage set of the decision boundary $\wout^\top \bm{x}_k = 0$, leading to their collision at $\rhohat=\rhohat_c$.

When $\rhohat>\rhohat_c$, intermittency is observed as the alternating sign of $z$, and the frequency increases with $\rhohat$ (Fig.~\ref{fig:fp09}(b)).
The mean span of sign reversals (i.e., the average residence time in attractor ruin) $\mean{\tau}$ decreases with a power law to $\rhohat-\rhohat_c$ (Fig.~\ref{fig:fp09}(c)). 
This power law, $\mean{\tau}\propto(\rhohat-\rhohat_c)^{-\gamma}$, is a characteristic of crisis-induced intermittency~\cite{grebogi1987critical, tanaka2005crisis}.

We analyzed the power spectral density (PSD) of a $10^6$-step time series of $z$.
For the distribution $P(f)$, we focused on low-frequency components ($f < 10^{-4}$) and estimated the slope on a log-log scale using linear regression: $\ln P(f) \sim \alpha \ln f + C$.
The results for $\hat{\rho} = 1.56$ and $\hat{\rho} = 1.80$ are shown in Figs.\ref{fig:fp09}(d)-(i) and \ref{fig:fp09}(d)-(ii), respectively.
As $\rhohat$ increases, $|\alpha|$ decreases (Fig.\ref{fig:fp09}(d)-(iii)).
Notably, near $\rhohat_c$, the low-frequency spectrum is well approximated by $f^{-1}$, and around $\rhohat = 1.9$, the slope nearly flattens.
This suggests that destabilization weakens trajectory autocorrelation, leading to a continuous transition from intermittency to fully merged chaos.
\FigureOne

The same bifurcation occurs regardless of the training point $\rho$ as long as $\Lambda^{\pm}$ are fixed points.
However, $\rho$ influences both $\rhohat_c$ and $\gamma$ by determining the distribution of the preimages of the decision boundary (SM~\four).

\FigureTwo
Next, the constant input is extended to a time series $\{\bm{u}_k\}$ by constructing the switching-driven system:
\begin{align}
    &\bm{z}_{k} = \Wout^\top \bm{x}_{k},~
    e^{\pm}_k = ||\pm\bm{u}_k-\bm{z}_k||,\\
    &\bm{x}_{k+1} = 
    \begin{cases}
        \R(\bm{x}_k, \bm{u}_k) & (e^{+}_k < e^{-}_k),\\
        \R(\bm{x}_k,  \bm{0}) & (e^{+}_k = e^{-}_k),\\
        \R(\bm{x}_k, -\bm{u}_k) & (e^{+}_k > e^{-}_k),
    \end{cases}
    \label{eq:switchingdriven}
\end{align}
where $\Wout$ is trained for the one-step-ahead prediction. 

Now, a three-dimensional trajectory of the Lorenz system is used as $\{\bm{u}_k\}$, focusing on the bifurcation induced by $\rhohat$.
Let $\bm{z}_k = [X_k, Y_k, Z_k]^\top$.
The readout is trained with the appropriate $\rho$, such that $e^{+}_k < e^{-}_k$ always holds for $\rhohat=\rho$.
Here, the parameters are set to $(\rho, \sigma)=(1.0, 0.025)$.
The attractor at $\rhohat=\rho$ corresponds to the response attractor $\Lambda^\pm$ driven by $\{\pm\bm{u}_k\}$.
As $\rhohat$ increases, the attractors expand, as seen in input-driven ESNs, and eventually collide (Fig.~\ref{fig:Lorenz}(a)-(i)).
Consequently, $\bm{z}_k$ transitions irregularly between the two attractor ruins, which maintain the shape of the original Lorenz attractors to some extent (Figs.~\ref{fig:Lorenz}(a)-(ii) and \ref{fig:Lorenz}(a)-(iii)).
The dynamics of this system is dominated by intermittent fluctuations in the prediction accuracy of $\Wout$.
Two factors likely contribute to this phenomenon: the robustness of $\Wout$ and the robustness of generalized synchronization, which breaks down accompanied by intermittent desynchronization~\cite{moskalenko2024intermittent, hramov2005intermittent, zhao2005transition}.

By feeding back $\bm{z}_k$ directly, this model becomes fully closed loop:
\begin{equation}
    \Rhat(\bm{x}) = (1-a)\bm{x} + a \tanh([\rhohat \W + \sigma \Win\Wout^\top] \bm{x})
    \label{eq:closedlooplesn}
\end{equation}
which reconstructs the Lorenz attractor (Fig.~\ref{fig:Lorenz}(b)-(i)). 
Increasing $\rhohat$ again induces an AMC accompanied by intermittency~\cite{flynn2023theory} (Fig.~\ref{fig:Lorenz}(b)-(ii) and \ref{fig:Lorenz}(b)-(iii)). 
We refer to this type of intermittency as ``double Lorenz."

Let $\drho=\rhohat-\rho$ and $\drho_c=\rhohat_c-\rho$.
Based on the time-series observation for $10^7$ steps, the estimated critical point is $-7\times10^{-8}<\drho_c<-6\times10^{-8}$. 
Therefore, the crisis has already occurred at $\rhohat=\rho$.
The statistical properties of output $Z$ are consistent; the slope $\alpha$ of the low-frequency PSD continuously changes from negative to zero, and the average residence time $\mean{\tau}$ decreases as a power law to $\rhohat-\rhohat_c$ (Fig.~\ref{fig:Lorenz}(b)-(iv) and \ref{fig:Lorenz}(b)-(v)).

In contrast to systems such as \eqref{eq:binaryfeedback} and \eqref{eq:switchingdriven}, $\rhohat$ directly affects the feedback value $\bm{z}$ in \eqref{eq:closedlooplesn}. 
Thus, understanding the emergence of the double-Lorenz attractor requires analyzing it as an autonomous system.

Figures \ref{fig:Lorenz_crisis}(a)-(i) and \ref{fig:Lorenz_crisis}(a)-(ii) show the inner activity $\bm{x}_k$ for $\drho=-10^{-7}$ and $\drho = 10^{-3}$, respectively, which are projected onto the space spanned by the first three principal components (PC1--PC3), obtained at $\drho=-10^{-7}$, which we call ``PC space."
The \Poincare section at $\mathrm{PC}1 = 10$ shows parabolical curve (Fig.~\ref{fig:Lorenz_crisis}(b)-(i)).
Trajectories are initially attracted to one of the four lobes of a pair of Lorenz attractors, distinguished by the signs of $Z$ and $Y$.
From this perspective, the state space can be divided into four regions, whose configuration on the $\mathrm{PC}1 = 0$ plane is shown in Fig.~\ref{fig:Lorenz_crisis}(b)-(ii).
At the region boundaries, two pairs of saddles, $P^{\pm}$ and $Q^{\pm}$, exist.
The points $P^{\pm}$ are located at the junctions of the four regions, while $Q^{\pm}$ lie along the $Z$-sign boundary, visible at the left and right edges of Fig.~\ref{fig:Lorenz_crisis}(b)-(iii).
The upper panel of Fig.~\ref{fig:Lorenz_crisis}(c) depicts trajectory fragments with $\mathrm{PC}1 > 0$ after the AMC, where the flow around $Q^+$ is observed.
The lower panels present multiple \Poincare sections along PC1, with each point clustered according to its near-future state, similar to Fig.~\ref{fig:Lorenz_crisis}(b)-(ii).
An escape pocket from the attractor ruin is identified at the tip of the surface, which emerges from the contact between the attractor and the stable manifold of $P^{\pm}$.
A schematic summary of the flow before and after AMC is presented in Fig.~\ref{fig:Lorenz_crisis}(d) (see also SM~\five).
This case study elucidated the details of a bifurcation scenario in which an acquired attractor comes into contact with a basin boundary that is intrinsically formed.
\FigureThree

The bistability of symmetric chaotic attractors realized by system \eqref{eq:closedloop} generally has the potential to induce intermittency via AMCs.
In particular, AMCs are closely related to chaotic itinerancy (CI)~\cite{ikeda1989maxwell, kaneko1990clustering, tsuda1991chaotic, komuro1999mechanism, tanaka2005crisis}, and these results suggest the possibility of designing itinerant dynamics through machine learning.

Since the induction of AMC in RNNs without feedback error utilizes the general properties of ESN, considering $\rhohat$ as the control parameter is reasonable.
While the emergence of the double-Lorenz attractor in this setup is somewhat universal---observed in all 200 ESN with different random realizations (SM  \five)---the induction of crisis by increasing $\rhohat$ is not necessarily guaranteed.
It may lead to other bifurcations, such as saddle-node or period-doubling bifurcations. 
If untrained attractors other than the targets exist, they can also be influential.
Thus, when utilizing the bifurcations of a trained RNN as a method for designing intermittency, it is essential to consider this uncertainity. 
We propose two approaches to address this issue.
One approach is to explore other bifurcation parameters, such as 
 the $(\rho, \sigma)$ plane. 
The other involves intervention at the training stage---specifically ensuring the robustness of $\Wout$ so that the emergence of untrained periodic orbits is avoided.
Therefore, potential countermeasures include introducing regularization (Method 1) or training $\Wout$ across multiple values of $\rho$ as follows (Method 2):
\begin{equation}
    [\bm{X}(\rho);~\alpha \bm{X}(\rho')] \Wout \approx [\bm{Y};~\alpha \bm{Y}].
\end{equation}
Method 2 can be extended into a parameter-aware approach~\cite{kim2021teaching, kong2021machine, kong2021emergence} by setting the training data $Y$ to the trajectories $Y'$.

Based on the above considerations, we examine several cases of AMC induction in RNNs.  
First, we consider an artificial case in which the governing equation of the target chaotic system is not explicitly defined.  
In RC-based embedding, training a periodic orbit with an excessively large $\rho$ can form a chaotic attractor along the periodic orbit~\cite{kabayama2025designing}.
Figures. \ref{fig:example}(a)-(i) and \ref{fig:example}(a)-(ii) depict a chaotic attractor formed along a Lissajous curve and crisis-induced intermittency induced by destabilization, respectively.
A similar phenomenon can be observed even when attractors overlap in the output space; Fig.~\ref{fig:example}(b) shows an ``at"-shaped chaotic attractor, constructed using a handwritten pattern, and the obtained intermittency. 
Here, the parameter is set to $p=1.0$.

Next, we explore applying this method to estimate AMC in target systems only by showing a single trajectory.
Specifically, we use Chua's circuit~\cite{chua1986double, sordi2020chua}, which has an AMC scenario whereby two symmetric spiral attractors merge to form a double-scroll attractor.
Here, a time series of a single spiral attractor is learned with Method 2; $(\rho, \rho', \alpha)=(1.5, 2.0, 0.25)$. 
At $\rhohat=1.5$, spiral attractors are observed (Fig.~\ref{fig:example}(c)-(i)). 
By increasing $\rhohat$, the double-scroll attractor is successfully induced (Fig.~\ref{fig:example}(c)-(ii)).
Even without Method 2---training with only one $\rho$, the birth of a double-scroll attractor can be observed in the RC model, but it is induced by other parameter manipulations (SM~\six).

We also consider a case with an asymmetric combination of attractors.
Figure~\ref{fig:example}(d)-(i) illustrates the case where a pair of Lorenz attractors is embedded using the MFRC with Method 2; $(\rho, \rho', \alpha) = (1.3, 1.31, 10^{-7})$ (SM~\seven).
Due to symmetry, each attractor has a corresponding twin, making the system four-stable.
These two pairs of attractors undergo crises and form a ``quadruple Lorenz'' attractor (Fig.~\ref{fig:example}(d)-(ii)).
\FigureFour

We compare our approach with the existing design method for intermittent dynamics using RC by \citet{inoue2020designing}. 
In~\cite{inoue2020designing}, artificially designed CI can be constructed by specifying quasi-attractors and probabilistic transition rules between them. 
While this method offers high controllability, it requires pretraining of the internal matrices of the RNN with online learning~\cite{laje2013robust}.
In contrast, our approach only requires embedding precursors of attractor ruins. 
While parameter exploration is necessary, the computational cost of a single trial remains low. 
Furthermore, due to the general properties of crises, the average residence time in the attractor ruin can be continuously adjusted by fine-tuning.
This kind of controllability was not available in~\cite{inoue2020designing}.
Additionally, a more reliable method based on this study is proposed in SM~\eight.

This paper analyzed several cases of AMCs and intermittency in RNNs.
For further details on RNNs in different settings, refer to SM~\three.

The manifestation of AMCs as intrinsic bifurcation scenarios in RNNs is also of interest from the perspective of neuroscience. 
\citet{flynn2023theory} utilized bifurcations in MFRC to construct a mathematical model of epilepsy from EEG time series. 
\citet{hadaeghi2013does, hadaeghi2015towards} proposed a mathematical model for bipolar disorder based on crisis-induced intermittency. 
In the context of a constructive approach to neural dynamics, exploiting crisis-induced intermittency in RNNs may offer a promising option.

\vskip\baselineskip
\begin{acknowledgments}
    K.A. was supported by JST Moonshot R\&D Grant No.~JPMJMS2021, the Institute of AI and Beyond of UTokyo, the International Research Center for Neurointelligence (WPI-IRCN) at The University of Tokyo Institutes for Advanced Study (UTIAS), JSPS KAKENHI Grant No.~JP20H05921, Cross-ministerial Strategic Innovation Promotion Program (SIP), the 3rd period of the SIP Grant No.~JPJ012207 and No.~JPJ012425.
    K.N. was supported by JSPS KAKENHI Grant No.~21KK0182 and 23K18472, by JST CREST Grant No.~JPMJCR2014, and by SIP on “Integrated Health Care System” Grant No.~JPJ012425.
\end{acknowledgments}

\bibliography{reference}
\end{document}


\title{
    Supplementary Material for\\
    ``Attractor-merging Crisis and Intermittency in Reservoir Computing''
}

\author{Tempei Kabayama}
\affiliation{
 Graduate School of Information Science and Technology, University of Tokyo, Bunkyo-ku, Tokyo 113-0033, Japan
}
\author{Motomasa Komuro}
\affiliation{
 International Research Center for Neurointelligence, University of Tokyo Institutes for Advanced Study, University of Tokyo, Bunkyo-ku, Tokyo 113-0033, Japan
}
\author{Yasuo Kuniyoshi}
\affiliation{
 Graduate School of Information Science and Technology, University of Tokyo, Bunkyo-ku, Tokyo 113-0033, Japan
}
\author{Kazuyuki Aihara}
\affiliation{
 International Research Center for Neurointelligence, University of Tokyo Institutes for Advanced Study, University of Tokyo, Bunkyo-ku, Tokyo 113-0033, Japan
}
\author{Kohei Nakajima}
\affiliation{
 Graduate School of Information Science and Technology, University of Tokyo, Bunkyo-ku, Tokyo 113-0033, Japan
}
\affiliation{
 International Research Center for Neurointelligence, University of Tokyo Institutes for Advanced Study, University of Tokyo, Bunkyo-ku, Tokyo 113-0033, Japan
}

\date{\today}
\maketitle

This Supplementary Material includes the following content:
\begin{itemize}
    \item Training methods of reservoir computing (RC)
    \item Setup of principal component analysis (PCA)
    \item Attractor-merging crisis (AMC) in random neural networks (RNNs) without training
    \item AMC in binary-feedback systems
    \item Birth and death of Double Lorenz
    \item Double scroll in RC
    \item Quadruple-Lorenz attractor
    \item Idea of a generic intermittency design method using a switching map
    \item References 
\end{itemize}



\newpage

\section{Training Methods of Reservoir Computing}
In this section,  we overview of the training method for RNNs using RC~\cite{jaeger2001echo,maass2002real,nakajima2021reservoir}, especially for the attractor reconstruction task. 
This approach employs an RNN referred to as a ``reservoir,'' which has a $D$-dimensional input $\bm{u}_k$ and an $N$-dimensional state $\bm{x}_{k}$, to perform one-step-ahead prediction of the input time series \eqref{supp_eq:rc_ol}.
By feeding back the predicted value $\bm{z}_{k} \approx \bm{u}_k$ as alternative input, an autonomous system is constructed \eqref{supp_eq:rc_cl}.
\begin{align}
    \bm{x}_{k+1}&=\R(\bm{x}_{k}, \bm{u}_k),~
    \bm{z}_{k}=\phi(\Wout^\top\psi(\bm{x}_k))\approx\bm{u}_k,\label{supp_eq:rc_ol}\\
    \bm{x}_{k+1}&=\R(\bm{x}_{k}, \bm{z}_{k})
                =\Rhat(\bm{x}_{k})\label{supp_eq:rc_cl}
\end{align}
The readout layer $\bm{W}_{out}\in\mathbb{R}^{N \times D}$ is trained using the following process:
The reservoir is driven by the teacher input (teacher forcing), and the set of the response is obtained for $T_{\mathrm{init}}+T_{\mathrm{train}}$ steps, with the initial $T_{\mathrm{init}}$ steps discarded. 
This yields the paired learning data:
\begin{equation}
        \bm{X} = \left[\psi(\bm{x}_{T_{\mathrm{init}}}) \cdots \psi(\bm{x}_{T_{\mathrm{init}}+T_{\mathrm{train}}})\right]^\top,
        \bm{Y} = \left[\bm{u}_{T_{\mathrm{init}}} \cdots \bm{u}_{T_{\mathrm{init}}+T_{\mathrm{train}}}\right]^\top.
\end{equation}
When $\phi$ is the identity function, $\Wout$ is determined by ridge regression or linear regression:
\begin{equation}
    \Wout = (\bm{X}^{\top}\bm{X} + \beta \bm{I})^{-1}\bm{X}^{\top}\bm{Y},
\end{equation}
where $\beta$ is the regularization parameter.

When $\phi$ is a sign function and $D=1$, the readout layer $\bm{w}_{out}\in\mathbb{R}^{N}$ can be trained, for example, using a linear support vector machine (SVM):
\begin{equation}
    \wout = \arg \min_{\wout} \frac{1}{2}||\wout||^2 ~\mathrm{s.t.}~ y_k(\wout^\top\psi(\bm{x}_k)) \geq 1.
\end{equation}

The output $\bm{z}$ is a low-dimensional projection of inner activity $\bm{x}$. Thus, we refer to $\bm{z}$ as the ``output space'' plot.

\section{Principal Component Analysis}
The method of principal component analysis used in this paper is outlined below. 
For an $N$-dimensional RNN, let $\bm{X}\in\mathbb{R}^{T\times N}$ be the matrix storing each state $\bm{x}_k$ along a trajectory of $T$ steps.  
To preserve symmetry concerning the origin, the principal component weights $\bm{p}_i$ are obtained from the covariance with a mean of $\bm{0}$ as follows:
\begin{align}
    &\bm{X}^{\top}\bm{X} = \bm{P}\bm{\Lambda}\bm{P}^{\top}\\
    &\bm{\Lambda} = \mathrm{diag}(\lambda_1, \ldots, \lambda_N), \quad \lambda_1 \geq \lambda_2 \geq \dots \geq \lambda_N,\\
    &\bm{P} = \left[\bm{p}_1, \ldots, \bm{p}_N\right]^\top, \quad ||\bm{p}_i|| = 1
\end{align}
The $i$-th principal component (PC) of the state $\bm{x}$ is given by  
$\mathrm{PC}i = \bm{p}_i^{\top} \bm{x}$. 
The projection onto the first three PCs $[\mathrm{PC}1, \mathrm{PC}2, \mathrm{PC}3]$ is referred to as the ``PC space'' plot.

\newpage

\section{Attractor-merging Crisis in Random Neural Networks without Training}
\subsection{Input-free RNN}
Consider the following RNN:
\begin{equation}
    \bm{x}_{k+1} = \mathcal{R}(\bm{x}_k)
    = (1-a)\bm{x} + a \tanh\left(\rho \W \bm{x}_k\right),
    \label{supp_eq:rnn}
\end{equation}
where $\bm{W} \in \mathbb{R}^{N \times N}$ is a random matrix whose elements follow a normal distribution, with a sparsity ratio of $p$ (i.e., the fraction of nonzero elements). The matrix is scaled to have a unity spectral radius.
Thus, $\rho$ corresponds to the spectral radius of the internal coupling matrix $\rho \bm{W}$. The parameter $a$ represents the leaky rate.  
This system has the origin as a stable fixed point for $\rho < 1$.
For sufficiently large $N$, it is known that this type of RNN exhibits chaos when $\rho > 1$~\cite{sompolinsky1988chaos, cessac2007neuron, jaeger2007optimization}.
An example of bifurcation for $(N, p, a) = (1000, 1.0, 0.5)$ is shown in Fig.~\ref{supp_fig:RhoBifVanilla}(a).
Here, the origin destabilizes through a supercritical Neimark-Sacker (NS) bifurcation, leading to a periodic orbit and eventually a chaotic attractor.  
In high-dimensional cases, the eigenvalue with the largest absolute value is likely to be complex, making this bifurcation pathway common.
However, if the corresponding eigenvalue is real, a pitchfork bifurcation occurs, resulting in two stable fixed points around the origin.  
An example of this scenario is shown in Fig.~\ref{supp_fig:RhoBifVanilla}(b). 
In this system, the two fixed points generated by the pitchfork bifurcation develop into chaotic states, and around $\rho = 1.55$, they undergo an AMC.
\RhoBifVanilla

\subsection{Periodically driven RNN}
AMC can also be observed in an RNN driven by external input.  
As an example, consider a case where the RNN \eqref{supp_eq:rnn} is driven by a sinusoidal wave:
\begin{align}
    \bm{x}_{k+1} &= \mathcal{R}(\bm{x}_k, u_k)
        = (1-a)\bm{x}_k + a \tanh\left(\rho \W \bm{x}_k + \sigma \win u_k\right)\label{supp_eq:periodic_driven}\\
    u_k &= \sin{\frac{\pi}{t}(k-t_0)}.
\end{align}
In this system, the symmetry  
$\R(-\bm{x}, u_{k+t}) = -\R(\bm{x}, u_k)$ 
holds. 
Thus, multistability can emerge, consisting of symmetric attractors: depending on the initial condition $(\bm{x}_0; t_0)$, the trajectory is attracted to different attractors.  
We set $\bm{x}_0$ at the origin for simplicity.  
Figure~\ref{supp_fig:driven}(a) visualizes the bistable state in terms of principal components for $(\rho, \sigma, t) = (1.16, 0.01, 50)$.  
Each attractor corresponds to $t_0 = 0$ and $t_0 = t$.  
As $\rho$ increases, this bistability collapses at a point, leading to intermittency as shown in Fig.~\ref{supp_fig:driven}(b).  
This phenomenon can be interpreted as analogous to the ``symmetry recovery'' observed in the Duffing equation~\cite{ishii1986breakdown}.
\RhoBifDriven

\section{Attractor-merging Crisis in Binary Feedback System}
The binary feedback system is described as follows:
\begin{align}
        \bm{x}_{k+1} &= \Rhat(\bm{x}_k)
        = (1-a)\bm{x}_k + a \tanh\left(\rhohat \W \bm{x}_k + \sigma \win \sgn(\wout^{\top}\bm{x}_k)\right)\label{supp_eq:binary_feedback_1}\\
        & = \begin{cases}
        \R^{+}(\bm{x}_k) & (\wout^{\top}\bm{x}_k > 0)\\
        \R^{0}(\bm{x}_k) & (\wout^{\top}\bm{x}_k = 0)\\
        \R^{-}(\bm{x}_k) & (\wout^{\top}\bm{x}_k < 0)
    \end{cases}\label{supp_eq:binary_feedback_2},
\end{align}
where
\begin{align}
    \R^{0}(\bm{x}) &= (1-a)\bm{x} + a \tanh\left(\rho \W \bm{x}\right)\label{supp_eq:no_input}\\
    \R^{\pm}(\bm{x}) &= (1-a)\bm{x} + a \tanh\left(\rho \W \bm{x} \pm \sigma \win\right).\label{supp_eq:constant_driven}
\end{align}
As $\rhohat$ increases, the attractors $\Lambda^{\pm}$ of the dynamical system $\R^{\pm}(\bm{x})$ evolve from fixed points to chaos, and these chaotic attractors expand their regions. 
Figure~\ref{supp_fig:binaryrho}(a) shows the bifurcation diagram of the system $\R^{+}$.

In the binary feedback system, $\wout$ is introduced as a linear separator for the two symmetric sets $\Lambda^{+}$ and $\Lambda^{-}$ using the linear SVM method.  
In particular, we focus on the case where $\Lambda^{\pm}$ are fixed points $\bm{p}^{\pm}$ at the trainnig point $\rhohat=\rho$.
In this case, $\wout$ is given by $\wout=\wout(\rho)=\bm{p}/\|\bm{p}\|^2$ and the system has at least two fixed points at $\rhohat=\rho$.
As $\rhohat$ increases, these fixed points develop into chaos, following a similar transition as in the systems $\R^{\pm}$.
At a certain point $\rhohat_c$, these chaotic attractors collide with the set $M$, which is the set of preimages of the decision boundary $\wout^{\top}\bm{x}=0$, leading to a crisis. 

The set $M$ at each $\rhohat$ is determined by the training point $\rho$. Thus, the critical point $\rhohat_c$ and the behavior after the crisis are also governed by $\rho$. 
Here, we analyze the relationship between $\rho$ and the set $M$.  
Since directly identifying $M$ is difficult, we introduce an approximate approach by considering the set $S$ of ``switching points'' defined as follows:
\begin{equation}
    \bm{x}^s \in S~\mathrm{if}~(\wout^{\top}\bm{x}^s)\cdot(\wout^{\top}\Rhat(\bm{x}^s)) \leq 0.
\end{equation}
Figure~\ref{supp_fig:binaryrho}(b)-(i) shows the proportion of switching points among the points randomly sampled from a uniform distribution $[-1, 1]$, for each $(\rho, \rhohat)$.  
As indicated by the red line, the proportion of switching points at the training point $\rhohat=\rho$ increases as $\rho$ becomes larger.
Furthermore, the rate of increase in this proportion when $\rhohat$ is increased is also larger for greater values of $\rho$.

The quantitative differences in $S$ that depend on the hyperparameter $\rho$ influence the decay of the mean residence time $\mean{\tau}$.  
Figure~\ref{supp_fig:binaryrho}(b)-(ii) shows $\log\mean{\tau}$ for each $(\rho, \rhohat)$, along with a fitted curve based on the power law approximation $\mean{\tau}=a\rho^{-\gamma} + b$.  
For the same $\rhohat$, the scale $\mean{\tau}$ becomes smaller as $\rho$ increases.
\BinaryRho

\section{Birth and Death of Double Lorenz}
\subsection{Embedding of Lorenz attractor}
The training data is obtained from the Lorenz system:
\begin{equation}
    \begin{aligned}
        \dot{x}&=10(y-x)\\
        \dot{y}&=x(28-z)-y\\
        \dot{z}&=xy-\frac{8}{3}z.
    \end{aligned}
    \label{supp_eq:Lorenz}
\end{equation}
We employ the echo state network (ESN) as follows:
\begin{equation}
    \bm{x}_{k+1} = (1-a)\bm{x}_k + a \tanh\left(\rho \W \bm{x}_k + \sigma \Win \bm{u}_k \right),
    \label{supp_eq:open_loop}
\end{equation}
and attractor reconstruction process results in the following autonomous system:
\begin{equation}
    \bm{x}_{k+1} = (1-a)\bm{x}_k + a \tanh\left([\rho \W + \sigma \Win \Wout^{\top}]\bm{x}_k\right).
    \label{supp_eq:closed_loop}
\end{equation}
The training data consist of trajectories obtained by solving the differential equation \eqref{supp_eq:Lorenz} using the fourth-order Runge-Kutta method. In this experiment, the trajectories were computed with a time step of $dt = 0,01$.

Noting the invariance of the system under the transformation $(x, y)\mapsto(-x, -y)$, trajectories were computed for two initial conditions, $(10, 10, 10)$ and $(-10, -10, 10)$, and both of them are used as teacher data after the first 30,000 steps were discarded.
This was done to mitigate the impact of trajectory residence time bias around the two equilibrium points $C_{\pm}=(\pm\sqrt{72}, \pm\sqrt{72}, 27)$: if the stability of the two saddles embedded in the reservoir is asymmetric, this untrained asymmetry originates from the reservoir's properties rather than the teacher time series. 

In the experiment, the hyperparameters for training were set as follows:
\begin{equation}\label{eq:lorenz_param}
(N, p, a, \rho, \sigma, T_{\mathrm{init}}, T_{\mathrm{train}}, \beta) = (1000, 0.01, 0.5, 1.0, 0.025, 5000, 5000, 0).
\end{equation}

\subsection{Saddle Points and the Flow Around Them}
\LorenzUM
The trained system \eqref{supp_eq:closed_loop} has four saddle points $C^{\pm}_{\pm}$, which form the centers of the two lobes of the Lorenz-like attractor.
In addition, it has three pairs of saddle points $S^{\pm}$, $P^{\pm}$, and $Q^{\pm}$, which contribute to the flow in the phase space.
The eigenvalues of these saddles are shown in Fig.~\ref{supp_fig:LE}(a).  
The saddle $S^{\pm}$ has one unstable real eigenvalue and one pair of stable complex eigenvalues with magnitudes close to one, which become the flow's rate-limiting factor.
This point lies on the surface of the attractor. 

As $\rhohat$ increases, each attractor gets close to $P^{\pm}$.
Figure \ref{supp_fig:LE}(b) shows the distance between attractor and $P^{\pm}$ which is calucurated by obtaining minimum value of $\|x - x^{(P)}\|$ in $10^6$-step trajectory.

The saddle $P^{\pm}$ has two unstable real eigenvalues, and $Q^{\pm}$ has one unstable real eigenvalue. These points lie on the basin boundaries.
Figure~\ref{supp_fig:LE}(c) shows the configuration of each point in the PC space and the flow along the unstable directions, which is obtained by calculating short-term trajectories from initial points that are slightly offset along the corresponding eigenvectors.
The unstable manifold (UM) of $S^{+}$ (U-S) is drawn into the left and right lobes surrounding $C^{+}_{\pm}$ of the Lorenz-like attractor.
The UM of $Q^{+}$ (U-Q) is drawn into the lobes surrounding $C^{+}_{-}$ and $C^{-}_{+}$ of the upper and lower Lorenz-like attractors, respectively.
The first UM of $P^{+}$ (U-P1) approaches each of $S^{\pm}$.
The trajectory corresponding to the second unstable eigenvalue is drawn into different attractor regions due to small differences in the initial state along the first unstable direction (U-P2(1), (2)).
At this boundary, it is hypothesized that the flow is drawn into $Q^{\pm}$.

A schematic summary of these flows is shown in Fig.~\ref{supp_fig:LE}(c).
The crisis occurs when the attractor surface, which is formed by the stable manifold swirling around $S$, collides with the stable manifold of $P$, which corresponds to the basin boundary.

\DestabS
Near the critical point, the contribution of $S$ to the attractor dynamics is relatively minor.
Trajectories approaching $S$ follow the vortex flow associated with the second and third eigenvalues but, upon reaching a certain proximity to $P$, diverge along the unstable directions (corresponding to the first eigenvalue of $S$ and the second eigenvalue of $P$).
For convenience, we refer to these unstable directions as the ``normal direction'' and define the plane that is neutral about the normal direction as the ``neutral plane.''
The point $S$ is a spiral sink within the neutral plane while being unstable along the normal direction.

At $\rhohat=1.00000$, when constraining the trajectory within the neutral plane---suppressing divergence along the normal direction---and tracking the flow near $S$, we can indeed observe a vortex attracted to $S$.

The stability of $S$ within the neutral plane is gradually lost as $\rhohat$ increases.
Figure \ref{supp_fig:destab_S}(b) shows the second and third eigenvalues of $S$ for each $\rhohat$.
These complex conjugate eigenvalues cross the unit circle near $\rhohat=1.00105$, leading to the loss of stability in the neutral plane.
This transition corresponds to a Neimark-Sacker (NS) bifurcation.

To capture the changes in flow near the bifurcation point in more detail, we track the trajectories around $S$ within the neutral plane for $\rhohat\in[1.00100,~1.00105]$ using the same five initial conditions. 
The results are shown in Fig.~\ref{supp_fig:destab_S}(c), where trajectories approaching $S$ over time are blue, while those diverging from $S$ are red.
At $\rhohat=1.00100$, all five trajectories are slowly attracted to $S$.
At $\rhohat=1.00101$, the outermost trajectory begins to diverge from $S$, suggesting the emergence of a saddle invariant circle at the boundary between the blue (approaching) and red (diverging) trajectory groups.
This invariant circle is unstable not only in the normal direction but also in the radial direction.
At this point, the contribution of $S$ to the double Lorenz attractor can be considered negligible.
As $\rhohat$ increases further, this saddle invariant circle shrinks and completely disappears at $\rhohat=1.00105$, where all trajectories diverge from $S$. 
Thus, this NS bifurcation is identified as a subcritical NS bifurcation, which is schematically summarized in Fig.~\ref{supp_fig:destab_S}(c)-(vi)

\subsection{Death of Double Lorenz and Emergence of Periodic Orbit}
\LorenzDeath
When $\rhohat$ is increased further, the system continues to exhibit an increase in transition frequency between attractor ruins, and at a certain point, a periodic orbit appears, causing the double Lorenz attractor to disappear (Fig. \ref{supp_fig:LD}(a)).
The appearance of this periodic orbit is due to a saddle-node bifurcation.
Considering the \Poincare section at $\mathrm{PC}2=0$, let $T_k$ represent the PC1 coordinate at the $k$-th crossing point.
The correspondence between $T_k$ and $T_{k+2}$ at $\rhohat>1.0290$ is shown in Fig.~\ref{supp_fig:LD}(b).
This \Poincare map exhibits a ``channel'' near $T_k=T_{k+2}$ at $\rhohat=1.0290$, and the system demonstrates Pomeau-Manneville-type intermittency~\cite{ott2002chaos}.
This ``channel'' is converted to a periodic orbit due to the saddle-node bifurcation near $\rhohat=1.0304$.

\subsection{Universality}
\LorenzUniv
Finally, we verify the universality of the results.
Keeping the teacher data and hyperparameters fixed at \eqref{eq:lorenz_param}, we vary only the realization of the random matrices $\W$ and $\Win$ while performing training and destabilization.
The behavior during destabilization is observed at sample points $\drho=10^{-4}, 10^{-3}, 10^{-2}$, and this process is repeated for $200$ trials.
We measure the transition probability as the fraction of a sign flip in the $Z$-component in the time series for $10^5$ steps.
If this probability is nonzero, we can at least conclude that the bistability of the Lorenz-type attractors has collapsed. 
Additionally, if the attractor ruin remains somewhat ``Lorenz-type,'' the system can be classified as a double Lorenz attractor.
To quantify the geometric similarity between the attractor ruin and the original Lorenz attractor, we compare the probability density along the $Y$-axis.
As an approximate method, we divide the $Y$-axis into $32$ bins and compute the residence frequency distribution for each bin. 
Let $p$ be the distribution obtained from the training data (ground truth), and $q$ be the distribution obtained from the system output.
The similarity of $q$ to $p$ is evaluated using the Kullback-Leibler (KL) divergence, $D_{KL}(p||q)$.
Here, we define the trajectory as similar to the original attractor if $D_{KL}(p||q)<0.5$.
For example, in the system analyzed in the main text, the double Lorenz attractor at $\drho=0.001$ yields $D_{KL}\simeq0.2$, while a periodic orbit emerging at $\drho=0.04$ results in $D_{KL}\simeq8.2$.

First, as a result, all systems in this experiment have a Lorenz-type attractor at the training point $\rhohat=\rho=1.0$; the maximum value of $D_{KL}$ in $200$ trials is $0.036$.
Here, one or more transitions occur in $25$ cases, indicating $\rho<\rhohat_c$.
Besides, the double Lorenz attractor was confirmed in all $200$ trials at one or more of the three sample points.
The results of these points are summarized in a two-dimensional histogram, as shown in Fig.~\ref{supp_fig:LU}.
The system's quantitative properties at each $\rhohat$ are concentrated around the mean without significant dispersion. 
Therefore, as long as the learning conditions are the same, the system's behavior changes with increasing $\rhohat$ exhibit quantitatively similar trends even with variations in the random matrices.
This confirms that the bifurcation scenario observed in this study has a certain level of universality.

\newpage

\section{Double scroll in Reservoir Computing}

The following Chua's circuit-based dynamical system is used as the target~\cite{chua1986double, sordi2020chua}:
\begin{equation}
    \begin{aligned}
        \dot{x}&=9.0[y-x-(bx-0.428\tanh(2x))]\\
        \dot{y}&=x-y+z\\
        \dot{z}&=-15y-0.125z.
    \end{aligned}
    \label{supp_eq:Chua}
\end{equation}
In this system, by varying $b$, a transition from a bistable state consisting of symmetric spiral attractors to a double-scroll attractor via an AMC is observed.

Similar to the embedding of the Lorenz system, the trajectories obtained from numerical calculations are used to train the RC system. Here, the time scale is set to $dt=0.05$.

First, consider the learning of the spiral attractor obtained at the bistable state $b=-0.5$.
Here, Method 2 is not applied, and the hyperparameters are as follows:
\begin{equation}
    (N, p, a, \rho, \sigma, T_{\mathrm{init}}, T_{\mathrm{train}}, \beta)=(1000, 0.01, 0.5, 1.5, 1.0, 10000, 10000, 0.001).
\end{equation}
In this case, AMC is not observed when $\rhohat$ is increased. 
The bifurcation diagram concerning $\rhohat$ is shown in Fig.~\ref{supp_fig:S2D}(a).
As $\rhohat$ increases, the system undergoes a reverse period-doubling bifurcation, and the spiral attractor degenerates into a periodic orbit.
As $\rhohat$ is increased further, a new periodic orbit appears, which undergoes a period-doubling bifurcation, leading to chaos and the emergence of AMC.
Conversely, the appearance of the double-scroll attractor due to AMC is confirmed when $\rhohat$ is decreased from the training point.

It is also possible to focus on parameters other than $\rhohat$.
For example, let $(\rhohat, \sigmahat)=(c\rho, c\sigma)$.
The scalar $c$ corresponds to the gain of the nonlinear function $\tanh$.
In the bifurcation caused by $c$, as shown in Fig.~\ref{supp_fig:S2D}(b), the appearance of the double-scroll attractor due to AMC is confirmed in the increasing direction of $c$.
\SpiraltoDoubleScroll

Next, the double-scroll trajectory obtained at $b=-0.614$ is learned using the same hyperparameters.
The learning results are shown in Fig.~\ref{supp_fig:D2S}(a)-(i). 
The bifurcation of this system in the $(\rhohat, \sigmahat)$ plane is analyzed.
Figure~\ref{supp_fig:D2S}(b) presents a color map that classifies the points $(\rhohat, \sigmahat)$ based on their corresponding attractors: periodic or non-periodic.
For non-periodic orbits, the orbit is further categorized based on whether the symmetric pairs of attractors are linearly separable (LS) in the output space, that is, obviously bistable.
As shown in Fig.~\ref{supp_fig:D2S}(a)-(ii--iv), the $(\rhohat, \sigmahat)$ plane exhibits bifurcation phenomena, where the double-scroll decomposes into two attractors through AMC, and this bifurcation occurs through multiple pathways.
\DoubleScrolltoSpiral

\section{Quadruple Lorenz attractor}
\QLorenzPC
The training data are reused from the Lorenz system used in the previous learning.
After normalizing the scale of each dimension to have zero mean and unit variance, the whole data is offset using the vector $(\pm1, 0, 1)$, resulting in the training data sets $\{\bm{u}_k^{(A)}\}$ and $\{\bm{u}_k^{(B)}\}$.
These are used for learning via multifunctional reservoir computing~\cite{flynn2021multifunctionality}, with modifications to train $\Wout$ for multiple values of $\rho$ (Method 2).
The hyperparameters are as follows:
\begin{equation}
    (N, p, a, \rho, \rho', \sigma, T_{\mathrm{init}}, T_{\mathrm{train}}, \alpha, \beta)=(1000, 0.01, 0.5, 1.30, 1.31, 1.0, 5000, 5000, 10^{-7}, 10^{-7}).
\end{equation}
At $\rhohat=\rho=1.30$, the system is four-stable: there are two pairs of Lorenz-type attractors. At $\rhohat=\rho'=1.31$, AMC has already occurred, and these four attractors are merging.
Figure~\ref{supp_fig:QLorenzPC} shows the dynamics for $\rhohat\in\{1.30, 1.31, 1.32\}$ in the PC space defined by $\rhohat=1.30$.
The frequency of transitions between attractor ruins increases according to $\rhohat$.

\section{Idea of a Generic Intermittency Design Method Using a Switching Map}
\DesignIdea
We briefly describe the idea of designing intermittency based on a multi-stable system using the framework of a switching map~\cite{kaneko2001complex}.

For example, we use the trained RNN with two Lorenz attractors discussed in this paper ($\rhohat=1-10^{-7}$) as the backbone bistable system (Fig.~\ref{supp_fig:design}(a)).
Here, the feedback function $f^{(1)}$ obtained through the attractor reconstruction is defined as follows:
\begin{equation}
    f^{(1)}(\bm{x}) = \sigma\Win\Wout^{\top}\bm{x}.
\end{equation}

First, we define the points $\bm{x}^{(A)}$ and $\bm{x}^{(B)}$ corresponding to the ``exit'' and ``entry'' points, respectively, in the attractors $\Lambda^{\pm}$.
Here, we choose $\bm{x}^{(A)}$ and $\bm{x}^{(B)}$, as shown in Fig.~\ref{supp_fig:design}(a).
Then, we place fixed points at these locations.
The bias input to the ESN that makes $\bm{x}^{(A)}$ the fixed point is analytically determined as follows:
\begin{equation}
    \begin{aligned}
        \bm{x}^{(A)} &= (1-a)\bm{x}^{(A)} + a \tanh(\rho \W \bm{x}^{(A)} + \bm{b}^{(A)})\\
        \therefore \bm{b}^{(A)} &= \arctanh(\bm{x}^{(A)}) - \rho \W \bm{x}^{(A)}.
    \end{aligned}
\end{equation}
Similarly, the bias input $\bm{b}^{(B)}$ for $\bm{x}^{(B)}$ is determined.
Starting with the initial value $\bm{x}^{(A)}$, we consider continuously changing the input from $\bm{b}^{(A)}$ to $\bm{b}^{(B)}$ to move the trajectory towards $\bm{x}^{(B)}$. 
First, we define the paths from $\bm{b}^{(A)}$ to $\pm\bm{b}^{(B)}$ using the following B\'{e}zier curve:
\begin{align}
    \bm{b}^{(\pm)}(\alpha)=
    -(\bm{b}^{(A)}\pm\bm{b}^{(B)})\alpha^{2}
    \pm2\bm{b}^{(B)}\alpha + \bm{b}^{(A)},
\end{align}
where the following is held:
\begin{align}
    &\bm{b}^{(\pm)}(0)=\bm{b}^{(A)},\\
    &\bm{b}^{(\pm)}(1)=\pm\bm{b}^{(B)},\\
    &\frac{d\bm{b}^{(+)}}{d\alpha}(0) = -\frac{d\bm{b}^{(-)}}{d\alpha}(0).
\end{align}
Along this path, the input bias is smoothly changed as follows:
\begin{align}
    &\bm{b}^{(\pm)}_k = \bm{b}^{(\pm)}(\alpha(t_k)))\\
    &\dot{\alpha} = -K(\alpha^3 - \alpha)\\
    &\alpha(0) = 10^{-4}
\end{align}
The rate of change at each step can be adjusted using a constant $K$ and timescale $dt$.
In this case, we chose $(K, dt)=(20, 0.01)$, and the change is set to complete the move in approximately 100 steps, as shown in Fig.~\ref{supp_fig:design}(b).
By driving the reservoir at $\bm{x}^{(A)}$ with this sequence $\{\bm{b}^{(\pm)}_k\}$, the trajectory indeed moves to $\pm\bm{x}^{(B)}$.
The projection of this trajectory onto the output space via $\Wout$ is shown in Fig.~\ref{supp_fig:design}(a).

Next, the sequence $\{\bm{b}^{(\pm)}_k\}$ is embedded into the reservoir. 
Here, it is necessary to ensure that the reservoir in a closed-loop configuration, starting with an initial condition near $\bm{x}^{(A)}$, stably outputs $\{\bm{b}^{(\pm)}_k\}$.
This is equivalent to making $\bm{x}^{(A)}$ a saddle point with a one-dimensional unstable eigenvalue.
The available methods include online training through FORCE learning~\cite{sussillo2009generating} and teacher forcing with noise; in this case, we use the latter.
The reservoir is driven by the sequence $\{\bm{b}^{(\pm)}_k\}$ with small noise $\bm{\eta}_k \sim \mathcal{N}(0,~10^{-6})$ added, producing the response time series $\{\bm{x}_k\}$.
By iterating this process, 20 time series are sampled, and the weight $\Wpath$ is trained to map all of these to the vicinity of $\{\bm{b}^{(\pm)}_k\}$.
Then, the feedback function $f^{(2)}$ is obtained as follows:
\begin{equation}
    f^{(2)}(\bm{x}) = \Wpath\bm{x}.
\end{equation}
Figure~\ref{supp_fig:design}(a) shows the eigenvalues of $\bm{x}^{(A)}$ under the feedback of $f^{(2)}$:
\begin{equation}\label{eq:design_f2}
    \bm{x}_{k+1} = (1-a)\bm{x}_k + a \tanh([\rho \W + \Wpath] \bm{x}_k),
\end{equation}
where it is confirmed that $\bm{x}^{(A)}$ indeed behaves as a saddle point with a one-dimensional unstable direction.

Based on the above, to design intermittency, the system can be constructed as follows:
\begin{equation}\label{eq:design}
    \begin{aligned}
        \bm{x}_{k+1} &= (1-a)\bm{x}_k + a \tanh(\rho \W \bm{x}_k + f_k(\bm{x}_k))\\
        f_{k+1} &= \left\{
                    \begin{array}{ll}
                        f^{(2)} & \mathrm{if~} ||\bm{x}_k\pm\bm{x}^{(A)}||< \epsilon\\
                        f^{(1)} & \mathrm{else~if~} ||\bm{x}_k\pm\bm{x}^{(B)}||<\delta\\
                        f_k & \mathrm{otherwise}.
                    \end{array}
                  \right.
    \end{aligned}
\end{equation}
In this system, the trajectory appears to probabilistically pass through the vicinity of $\pm\bm{x}^{(A)}$.
When it gets sufficiently close, a switching of the feedback map occurs, causing the trajectory to escape the attractor.
The trajectory moves along a path toward $\pm\bm{x}^{(B)}$ or $\mp\bm{x}^{(B)}$ in phase space.
Once the transition is complete, another switching of the map occurs, and the trajectory remains in the attractor until the next switch.

The parameters $(\epsilon, \delta)$ should be taken carefully because they need to be sufficiently large to introduce variability at the points where the map switches. If they are too small, the switching may occur periodically.

\providecommand{\noopsort}[1]{}\providecommand{\singleletter}[1]{#1}%